\documentclass[epsfig,prd,aps,preprint,superscriptaddress,showpacs,showkeys]{revtex4}
\usepackage{epsfig}
\usepackage{graphics}

\begin{document}
\title{Asymptotic behavior of the warm inflation scenario with viscous pressure}
\author{Jos\'e P.~Mimoso}
\email{jpmimoso@cii.fc.ul.pt}
 \affiliation{Department of Physics, Faculdade de Ci\^encias da Universidade de Lisboa, \\
and Centro de F\'{\i}sica Te\'orica e Computacional da Universidade de Lisboa,\\
 Av. Prof. Gama Pinto 2,  P-1649-003 Lisboa, Portugal}
 \author{Ana Nunes}
\email{anunes@lmc.fc.ul.pt}
 \affiliation{Department of Physics, Faculdade de Ci\^encias da Universidade de Lisboa,\\
and Centro de F\'{\i}sica Te\'orica e Computacional da Universidade de Lisboa,\\
 Av. Prof. Gama Pinto 2,  P-1649-003 Lisboa, Portugal}
\author{Diego Pav\'on}
\email{diego.pavon@uab.es} \affiliation{Departamento de
F\'{\i}sica, Universidad Aut\'onoma de  Barcelona, Facultad de
Ciencias\\ 08193 Bellaterra (Barcelona) Spain}
\date{\today}

\begin{abstract}
We analyze the dynamics of models of warm inflation with general
dissipative effects. We consider phenomenological terms both for
the inflaton decay rate and for viscous effects within matter. We
provide a classification of the asymptotic behavior of these
models and show that the existence of a late-time scaling regime
depends not only on an asymptotic behavior of the scalar field
potential, but also on an appropriate asymptotic behavior of the
inflaton decay rate. There are scaling solutions whenever the
latter evolves to become proportional to the Hubble rate of
expansion regardless of the steepness of the scalar field
exponential potential. We show from thermodynamic arguments that
the scaling regime is associated to a power-law dependence of the
matter-radiation temperature on the scale factor, which allows a
mild variation of the temperature of the matter/radiation fluid.
We also show that the late time contribution of the dissipative
terms alleviates the depletion of matter, and increases the
duration of inflation.
\end{abstract}


\maketitle

\section{Introduction}\label{sec_introduction}
There is a widespread belief that our Universe, or  at least a
sufficiently large part of it causally connected to us,
experienced an early  period of accelerated expansion, called
inflation. This happened before the primordial nucleosynthesis era
could take place and likely after the Planck period. Many
inflationary scenarios have been proposed over the years~(see
\cite{Inflation} and references therein). Most of them rely on the
dynamics of a self-interacting scalar field (the ``inflaton''),
whose potential overwhelms all other forms of energy during the
relevant period. They generally share the unsatisfactory feature
of driving the Universe to such  a super-cooled state that it
becomes necessary to introduce an ad hoc mechanism -termed
``reheating"- in order to raise  the temperature of the universe
to levels compatible with primordial nucleosynthesis. Therefore
this reheating phase appears as a subsequent, separate stage~
mainly justified by the need to recover from the extreme effects
of inflation \cite{Berera Pascos} during which a rather elaborate process
of multifield parametric resonances followed by particle production ~\cite{Kofman et al
97,Bassett+Maartens 99,Charters+Nunes+Mimoso 05a} takes place.

As an alternative some authors have looked for inflationary
scenarios leading the universe to a moderate temperature state at
the end of the superluminal stage so that the reheating phase
could be dispensed with altogether. It was advocated that this can
be accomplished by coupling the inflaton to the matter fields in
such a way that the decrease in the energy density of the latter
during inflation is somewhat compensated by the decay of the
inflaton into radiation and particles with mass. This would happen
when the inflaton rolls down its potential, but keeping the
combined pressure of the inflaton and radiation negative enough to
have acceleration. This kind of scenario, known as ``warm
inflation" as the radiation temperature never drops dramatically,
was first proposed by Berera~\cite{Berera 95a,Berera 95b}. It now
rests on solid grounds since it has been forcefully argued in a
series of papers that indeed the inflaton can decay during the
slow-roll (see, e.g. \cite{Berera+Ramos 05,Hall,Mar} and
references therein). Besides, this scenario has other
advantages, namely: $(i)$ the slow-roll condition $\dot{\phi}^2 \ll%
V(\phi)$ can be fulfilled for steeper potentials, $(ii)$ the
density perturbations generated by thermal fluctuations may be
larger than those of quantum origin~\cite{origin,Hall+Moss+Berera
03,Gupta etal 02}, and $(iii)$ it may provide a very useful
mechanism for baryogenesis \cite{baryogenesis}.

To simplify the study of the dynamics of warm inflation, previous
works treated the particles created in the decay of the inflaton
purely as radiation, thereby ignoring the existence of particles
with mass in the decay fluid. Here, we will go a step beyond by
taking into account the presence of of particles with mass as part
of the decay products, and give a hydrodynamical description of
the mixture of massless and non-massless particles by an overall
fluid with equation of state $p = (\gamma- 1)\rho$, where the
adiabatic index $\gamma$ is bounded by $1 \leq \gamma \leq 2$.

On very general grounds, this fluid is expected to have a negative
dissipative pressure, $\Pi$, that somewhat quantifies the
departure of the fluid from thermodynamical equilibrium, which we
will consider to be small but still significant. This viscous
pressure arises quite naturally via two different mechanisms,
namely: $(i)$ the inter-particle interactions \cite{landau}, and
$(ii)$ the decay of particles within the matter fluid
\cite{Ya-JDB}.

A well known example of mechanism $(i)$ of
prime cosmological interest is the radiative fluid, a mixture of
massless and non-massless particles, as it plays an essential role
in the description of the matter-radiation decoupling in the
standard cosmological model
\cite{radiative1,radiative2,radiative3}.

A sizeable viscous pressure
also arises spontaneously in
mixtures of different particles species, or of the same species
but with different energies -a typical instance in laboratory
physics is the Maxwell-Boltzmann gas \cite{Harris}. One may think
of $\Pi$ as the internal ``friction" that sets in as a consequence
of the diverse cooling rates in the expanding mixture, something
to be expected in the matter fluid originated by the decay of the
inflaton.

As for mechanism $(ii)$, it is well known that the decay of
particles within a fluid can be formally described by a bulk
dissipative pressure $\Pi$. This is only natural because the decay
is an entropy--producing  scalar phenomenon associated with the
spontaneous enlargement of the phase space (we use the word
``scalar" in the sense of irreversible thermodynamics) and the
bulk viscous pressure is also a scalar entropy--producing agent.
There is an ample body of literature on the cosmological
applications of this analogy -see e.g. \cite{Ya-JDB},
\cite{analogy}. In the case of warm inflation, it is natural to
expect that, at least, some species of particles directly produced
by the decay of the inflaton will, in turn, decay into other,
lighter species. In this connection, it has been proposed that the
inflaton may first decay into a heavy boson $\chi$ which
subsequently decays in two light fermions $\psi_{d}$ \cite{decay}.
This is an obvious source of entropy, and therefore it can be
modelled by a dissipative bulk pressure $\Pi$.

Our purpose in this paper is to generalize the usual warm
inflationary scenario by introducing the novel elements mentioned
above, namely  the decay of the scalar field into a fluid of
adiabatic index $\gamma$ rather than just radiation, and
especially the dissipative pressure of this fluid, irrespective of
the underlying mechanism. We will not dwell on the difficult
question of the quantum, non--equilibrium thermodynamical problem
underlying warm inflation~\cite{Berera+Ramos 05,Calzetta+Hu
88,Yokoyama+Linde 99,Berera+Gleiser+Ramos 98,Berera+Gleiser+Ramos
99,Moss 02,Lawrie 02}, but rather take a phenomenological approach
similar to that considered in several works~\cite{Oliveira+Ramos
98,Bellini 98,Maia+Lima 99,Yokoyama+Sato+Kodama 87,Billyard+Coley
00} (which can be traced back to the early studies of
inflation~\cite{Albrecht et al 82}). Instead of adopting a model
building viewpoint and looking for the implications of specific
assumptions, we aim at identifying typical features of models that
yield interesting asymptotic behavior. We resort to a qualitative
analysis of the corresponding autonomous system of differential
equations using the approach developed in~\cite{NM 00} that allows
the consideration of arbitrary scalar field potentials. We will
characterize the implications of allowing for various forms of the
rate of decay of the scalar field, as well as for various forms
for the dissipative pressure. We consider, for instance, models
with scalar field potentials displaying an asymptotic exponential
behavior. These arise naturally in generalized theories of gravity
emerging in the low-energy limits of unification proposals such as
super-gravity theories or string theories \cite{exponential1}. On
the one hand, after the dimensional reduction to an effective
4-dimensional space-time and the subsequent representation of the
theories in the so-called Einstein frame typical polynomial
potentials become exponential~\cite{Wetterich 94,NMC 01}. On the
other hand, the theories are then characterised by the  existence
of a scalar field that couples to all non-radiation fields, with
the coupling depending, in general, on the scalar field. The
simplest example of these features can be found in the so-called
non-minimal coupling theories. We provide a classification of the
relevant global dynamical features of the cosmological model
associated with those possible choices. A limited account of some
of the results of the present work  was reported in \cite{MNP
AIP04}.

One question we address is whether {\it non-trivial} scaling
solutions~\cite{Ratra+Peebles 88,WCL 93,NM 00,Billyard+Coley 00}
(hereafter simply termed scaling solutions) exist, i.e., solutions
where the ratio of energies involving the matter fluid and scalar
field keep a constant ratio. Another class of solutions refered in the literature as having a
scaling asymptotic behavior are those for which both the energy
density of the scalar field and that of the matter fluid decay
with different power laws of the scale factor of the
universe~\cite{Ratra+Peebles 88,Zlatev et al 99,Liddle+Scherrer
99}. In this latter case one of the components eventually
dominates and thus the ratio of their energy densities becomes
evanescent, in clear contrast to the case of the non-trivial
scaling solutions. We shall term these solutions as trivial
scaling solutions to contrast them with the previous ones
sometimes dubbed tracker solutions. The trivial case arises in
association with scalar field potentials of a power-law type and,
as we shall see, they occur when the scalar field decays have the
the same type of time-dependences as those required by the (non-trivial,
tracking) scaling solutions.

One of the reasons why non-trivial scaling solutions are important
is that they provide an asymptotic stationary regime for the
energy transfer between the scalar field and radiation. This
stationary  (sometimes termed ``quasi-static'') regime is an
assumption in the standard treatment of warm
inflation~\cite{origin} to evaluate the temperature of matter in
the final stages. On the other hand, introducing this class of
solutions in the kinetic analysis of interacting
fluids~\cite{Zimdahl+Triginer+Pavon 96,Zimdahl+Pavon 01} leads to
an alternative to the usual $\Gamma \gg 3H$ case, generalizing the
example of Ref.~\cite{Lima+Carrillo} where temperature of the
matter (radiation) bath is nearly constant.

We show that this class of scaling behavior depends not
only on the asymptotic form of the inflaton~\cite{NM 00}, but also
on having an appropriate time-dependent rate for the scalar field
decay. The additional consideration of bulk viscosity, besides
being a natural ingredient in models with one or more matter
components as well as in models with inter-particle decays,
facilitates the Universe to have a late time de Sitter expansion.

An outline of this work is as follows. Section II studies the
model underlying the original idea of the warm inflation proposal,
namely the model in which the inflaton field decays into matter
during inflation thus avoiding the need for the post-inflationary
reheating. This decay is characterized by a rate $\Gamma$ which we
shall initially assume to be a constant. Our results though will
argue in favour of a varying $\Gamma$ and we shall thus consider
the case where $\Gamma \propto H$. This yields late time scaling
solutions whenever the scalar field potentials asymptotes to an
exponential behavior. This happens regardless of the slope of the
potential. Subsequently, section IIII, analyses more realistic
models where a bulk viscous pressure term $\Pi$ is also present in
the equation of state of matter. We first envisage the usual form
$\Pi= -3\zeta H$ for that pressure and, subsequently, analyse a
general model with both a varying rate of decay and a general form
for the bulk viscosity $\Pi= -3\zeta \, \rho^\alpha\,H^\beta$,
where $2\alpha+\beta=2$ on dimensional grounds. Finally, section
IV provides a discussion of our results.

\section{The dynamics of Warm inflation} \label{sec_dyn_warminf}
\subsection{Warm inflation with constant $\Gamma$}\label{Sec_Gammaconst}

We consider a  spatially flat Friedmann-Robertson-Walker universe
filled with a self-interacting scalar field and a perfect fluid
consisting of a mixture of matter and radiation, such that the
former decays into the latter at some constant rate $\Gamma$. For
the time being we ignore the dissipative pressure. We also neglect
radiative corrections to the inflaton
potential~\cite{Hall+Moss+Berera 03,Yokoyama+Linde 99}. The
corresponding system of equations reads
\begin{eqnarray}
3H^2 &=& \rho + \frac{\dot{\phi}^2}{2} + V(\phi) \, ,\label{eq_Fried}\\
\dot{H}&=& - \textstyle{1\over{2}}({\dot{\phi}^2+\gamma \rho})\, ,  \label{eq_Ray}\\
\ddot{\phi} &=& - (3H+\Gamma)\dot{\phi}-V'(\phi)\, ,
\label{eq_KG}\,
\end{eqnarray}
\\
where here and throughout we use units in which $8\pi G= c = 1$.
The first two are Einstein's equations, the third describes the
decay of the inflaton. From these, it follows the energy balance
for the matter fluid,
\\
\begin{equation}
\dot{\rho} = -3\gamma\,H\, \rho + \Gamma \dot{\phi}^2 \, .
\label{eq_cons}
\end{equation}
\\
As usual $H \equiv \dot{a}/{a}$ denotes the Hubble factor.

To cast the corresponding autonomous system of four differential
equations it is expedient to introduce the set of normalized variables
\begin{eqnarray}
x^2 &=& \frac{\dot{\phi}^2}{6H^2}\\
y^2 &=& \frac{V(\phi)}{3H^2}\\
r &=& \frac{\Gamma}{3H} \; ,
\end{eqnarray}
along with the new time variable $N= \ln{a}$. Thus we get
\begin{eqnarray}
x' &=& x \,\left( Q-3(1+r)\right)- W(\phi) \, y^2, \label{x'_1} \\
y' &=& \left(Q+W(\phi)\, x\right) y\; , \label{y'_1}\\
r' &=& r\,Q \; , \label{r´_1} \\
\phi' &=& \sqrt{6} \, x \; , \label{phi'_1}
\end{eqnarray}
where a prime means derivative with respect to $N$, and the
definitions
\begin{equation}
 W(\phi) =\sqrt{\frac{3}{2}}\, \left(\frac{\partial_{\phi}V}{V}\right)
\end{equation}
 and
\begin{equation}
Q =\frac{3}{2}\,\left[2x^2+\gamma\,
(1-x^2-y^2)\right]  \; , \label{defQ_1}
\end{equation}
as well as $\rho/(3H^2)=1-x^2 - y^2$ were used. Equation
(\ref{phi'_1}) was first considered in \cite{NM 00}, and is
crucial for the consideration of general potentials $V(\phi)$
besides the particular case of the exponential potential. The
function $Q$ defined by  Eq.~(\ref{defQ_1}) is related to the
deceleration parameter $q=-\ddot{a}a/\dot{a}^2$ by $Q=1+q$.

The special case where $r=0$, naturally, corresponds to the absence of
interaction between the scalar field and the perfect fluid, and it is an
invariant manifold of the dynamical system~(\ref{x'_1})-(\ref{phi'_1}).
It is appropriate to refer here its major features in order to better
appreciate the implications of the decay of the scalar field
(see Table \ref{table1}).

We distinguish the fixed points of the system into those occurring
for finite values of $\phi$ and those associated with the
asymptotic limit, $\phi\to \infty$. In the former case, i.e., for
finite $\phi=\phi_\ast$, the fixed points always require the
vanishing of the kinetic energy of the scalar field ($x=0$). They
are located at the origin ($x=y=0$), and at $(x=0,y=1)$, on the
frontier of the phase space domain $x^2+y^2=1$, which is an
invariant manifold. For $x=0$, $y=0$, the potential must have a
vanishing critical point at $\phi_\ast$, a case that cannot be
dealt with the variables in use, but it is well-know that if
$\phi_\ast$ is a minimum at the origin, then it is a stable point
and the scale factor evolves as $a(t) \propto t^{2/(3\gamma)}$
\cite{Belinski et al 85,Liddle+Scherrer 99}. The fixed points on
$x^2+y^2=1$ are given by $x=0$, $y=1$ and require that $W=0$. This
means that they can only occur in association with extrema of the
potential. Their stability is defined by the sign of
$W'(\phi_\ast)$, where $\phi_\ast$ is the value of $\phi$ where
$V'(\phi)$ (and hence $W$) vanishes. When $V$ has a non-vanishing
minimum and, hence $W'>0$, the critical point is a stable node.
When $V$ has a maximum and, hence $W'<0$, we have a saddle point
(an unstable fixed point). These fixed points correspond to the de
Sitter exponential behavior and are accompanied
 by the depletion of the matter component ($\rho =0$).

To study the critical points that occur at $\phi\to \infty$ (which
we shall label $\phi_\infty$), we carry out the regularization
produced by the change of variable $\psi= 1/\phi$. Then
Eq.~(\ref{phi'_1}) becomes
\begin{equation}
\psi'= - \sqrt{6}\, x \, \psi^2 \; , \label{phi'_1b}
\end{equation}
and the critical points correspond to either to $x=0$, as
previously seen, or $\psi=0$. The $\phi_\infty$ critical points
depend  on the asymptotic behavior of $V(\phi)$~\cite{NM 00}. If
$V(\phi)$ exhibits some non-vanishing  asymptotic value we have
again $x=0$, $y=1$ corresponding to a cosmological constant and,
hence, to a de Sitter late-time behavior. If $V(\phi)$ asymptotes
towards the exponential potential, say $V\propto e^{-\lambda
\phi}$, with $\lambda$ constant, there are several possible fixed
values dependent on the ratio between $\lambda^2$ and $\gamma$
(see, for instance,~\cite{CLW 98} for details).  There are
unstable fixed points on the invariant manifolds bounding the
phase-space domain for all possible choices of both
$W=-\sqrt{3/2}\,\lambda$ and $\gamma$, namely: $(i)$ a matter
dominated solution at $x=0$ and $y=0$, which is a saddle and
corresponds to $a(t) \propto t^{2/(3\gamma)}$, $(ii)$ two
solutions dominated by the scalar field kinetic energy at $x=\pm
1$ and $y=0$ which are  either unstable nodes or saddles, and
correspond to the stiff behavior $a(t) \propto t^{1/3}$,
$\phi_\infty(t)\sim \ln{t^{K_0}}$, where $K_0$ is an arbitrary
constant defining the scalar field initial velocity. There is
another fixed point on the $x^2+y^2=1$ boundary representing a
scalar field dominated solution, when $W^2<9$. This fixed point is
stable when $W^2<9\gamma/2$, and unstable otherwise (saddle). Thus
for $W^2>9\gamma/2$ (i.e., $\lambda^2 >3\gamma$), there is a
stable fixed point in the interior of the phase space domain.
This latter point  corresponds to scaling behavior between the
matter and scalar field energy-densities~\cite{Ratra+Peebles
88,Wetterich 88,WCL 93,Ferreira+Joyce 97,CLW 98,NMC 01}. This
attractor solution is characterized by $a(t) \propto
t^{2/3\gamma}$ and $\phi-\phi_0 = \ln{t^{\pm 2/\lambda}}$.

There are also trivial scaling solutions for which $\rho_\phi
\propto a^{-n}$ and $\rho\propto a^{-m}$, where $n> m$ are
positive constants, when~\cite{Liddle+Scherrer 99}
\begin{equation}
V(\phi) =A^2 \,\left(1-\frac{n}{m} \right)^2\,\left( \frac{6-n}{2n}\right)\,
\left(\frac{\phi}{A}\right)^\varpi
\label{tri_ssol_1}
\end{equation}
where
\begin{equation}
\varpi = \frac{2n}{n-m} \; . \label{tri_ssol_2}
\end{equation}

Coming back to the model that includes the interaction and thus
letting $r$ be non-vanishing, we immediately see from
Eq.~(\ref{r´_1}) that, along the $r$-direction, all the points are
singular points if and only if $Q=0$. For finite values of $\phi$,
as $x=0$ at the fixed points, this requires once more $y^2=1$ so
that the singular points are associated with $\rho = 0$, i.e.,
with the depletion of the matter component. Moreover, as in the
$r=0$ case, these singular points are extrema of the potential
$V(\phi)$. They correspond to a de Sitter behavior ($a(t)\propto
e^{\sqrt{V(\phi_0)/3})\, t}$, $\phi=\phi_0$ constant) and are
either stable or unstable depending on the extremum being a
minimum ($W'>0$) or a maximum ($W'<0$). In fact, at the singular
points corresponding to extrema of the potential $V(\phi)$, the
eigenvalues found in the linear stability analysis are
\begin{eqnarray}
\mu_y &=& -3\gamma \, , \label{vap_r_y}\\
\mu_{x,\phi}&=& -\frac{3(1+r)}{2} \left[1\pm \sqrt{1- \frac{4\sqrt{6} W'(\phi_0)}{9(1+r)^2} } \right]\; .
\label{vap_t_xphi}
\end{eqnarray}
On the other hand, we no longer have fixed points at  $x=0$, $y=0$ (unless $\gamma=0$ which corresponds
to the perfect fluid being a cosmological constant). This happens because the system then evolves along
the $r$-axis towards $r\to \infty$, a behavior that can only be prevented by the existence of a positive
minimum of the potential $V(\phi)$. At $\phi \to \infty$ the system does not exhibit scaling solutions
anymore. The only fixed points allowed in this asymptotic limit are those associated with a non-vanishing,
asymptotically flat potential, which thus corresponds to the de Sitter exponential behavior.

\begin{figure}
\begin{center}
\psfig{file=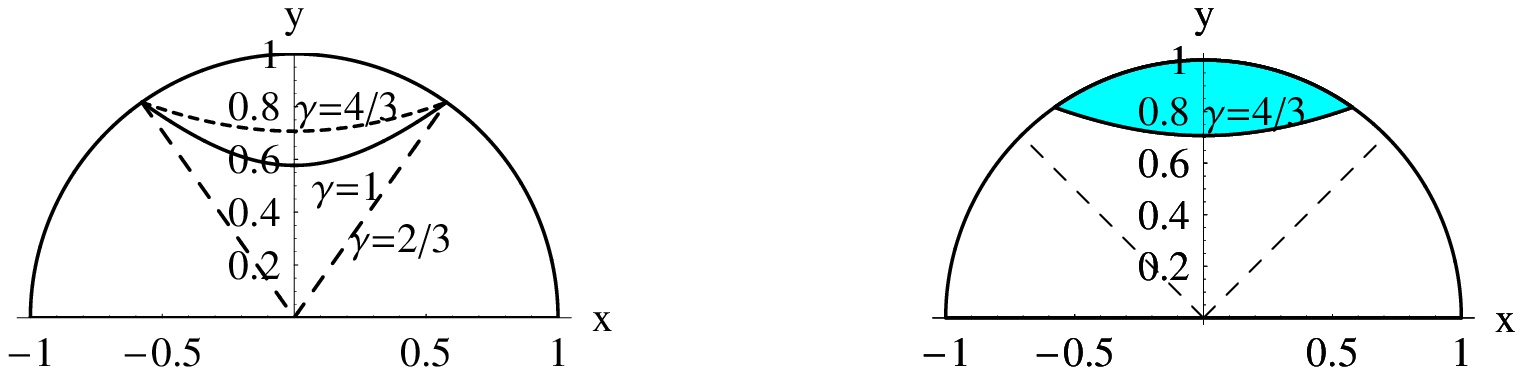,height=4cm,width=12cm}
\end{center}
\caption{Inflationary region of the models without bulk viscosity.
Inflation occurs in the shaded region between the hyperbole and
the boundary of the phase space. The figure on the left depicts
the variation of the region with $\gamma$, whereas in the figure
on the right we have taken $\gamma=4/3$.}
  \label{fig1}
\end{figure}

Accelerated expansion corresponds to the region of the phase space where $Q<1$, so that
\begin{equation}
3\gamma y^2 - 3(2-\gamma) x^2 > 3\gamma - 2 \; . \label{inflation}
\end{equation}
This condition does not carry any dependence either on $r$ or
$\phi$. Thus we may restrict our discussion to a $(x,y)$
projection of the phase space. The condition~(\ref{inflation})
defines for $1<\gamma<2$ the region between the upper branch of
the hyperbolae $3\gamma y^2 - 3(2-\gamma) x^2 = (3\gamma - 2)$ and
the boundary $x^2+y^2=1$ of the phase space domain (see
Figure~\ref{fig1}). The asymptotes of the hyperbolae are $y=\pm
\sqrt{(2-\gamma)/\gamma}$, and we see that, as $\gamma$ increases,
the inflationary region becomes progressively smaller. In fact the
region shrinks vertically towards the $x=0$, $y=1$ point and it
reduces to it in the limit case of $\gamma=2$.

In Ref.~\cite{origin} the end of inflation is given by the
condition $\rho_\phi\simeq\rho_\gamma$ and  this event is
associated with the beginning of the matter (radiation)
domination. As it becomes apparent from the above discussion, the
condition for the end of inflation, $Q = 1$, is more general and
does not strictly require matter domination. Taylor and Berera's
condition  \cite{origin} corresponds to the end of slow-roll
inflation (i.e., $\dot{\phi}^2\simeq 0\simeq x$) and is extended,
in the present study, to general $\gamma$-fluids  as
\begin{equation}
\rho_m \simeq \frac{2}{3\gamma-2}\,\rho_\phi \; .
\end{equation}

The independence on $r$ of the size of inflationary region should
not though be understood as the interaction having no effect on
inflation. From Eqs~(\ref{vap_r_y}) and (\ref{vap_t_xphi}) we see
that the eigenvalues of the linearized system at the fixed points
carry a dependence on $r$ which is such that it renders  the
minima of the potential more stable and the maxima less unstable
(as if the potential became shallower). Thus the transfer of
energy from the scalar field to the perfect fluid favors inflation
in that the system spends a longer time in the neighborhood of the
extrema of the potential. This is exactly what is meant to happen
in the warm inflation scenario where it is assumed that slow roll
holds and argued that $r$ allows for steeper potentials than those
required in its absence. As discussed in \cite{origin}, it is a
simple matter to see that the slow-roll condition on $\dot{\phi}$
\begin{equation}
\dot{\phi} = - \frac{V'}{3H(1+r)}\simeq -\frac{V'}{3rH} \; ,
\end{equation}
is easier to satisfy if the scalar field decays, that is, if $r>1$
and much easier if $r\gg 1$.

The fact that $r$ increases indefinitely in the present model is a
consequence of its definition, and merely translates the fact
that, unless the system is trapped at a non-vanishing minimum of
$V(\phi)$, $H$ decreases towards zero. Since this is a direct
result of assuming a constant $\Gamma$, we consider in the next
section a more appropriate model where $\Gamma$ decreases as the
Universe's expansion proceeds.

\subsection{Warm inflation with $\Gamma \propto H$}\label{Sect_rproptoH}
We assume that $\Gamma_\phi = 3\Gamma_\ast H$ where $\Gamma_\ast$
is a dimensionless, positive constant. As $H$ is expected to be a
non-increasing function of time in an expanding universe, this is
a simple choice for the time dependence of $\Gamma_\phi$ such that
the decays have a stage of maximum intensity (when inflation
occurs) followed by a progressive attenuation until it vanishes
altogether.

Now $r=\Gamma_\ast$ is a constant parameter and the dynamical system reduces
to the three equations
\begin{eqnarray}
x' &=& x \,\left[Q-3(1+r)\right]- W(\phi) \, y^2, \label{x'_2} \\
y' &=& \left[Q+W(\phi)\, x\right] y\; , \label{y'_2}\\
\phi' &=& \sqrt{6} \, x \; , \label{vphi'_2}
\end{eqnarray}
where $Q$ is still given by Eq.~(\ref{defQ_1}) We see that these
equations are analogous to those of the $r=0$ case of the previous
section with a different coefficient on the linear term in $x$ of
Eq.~(\ref{x'_2}). Thus the basic qualitative dynamical features
remain the same as those found for that model (see Table
\ref{table2}). The decay of the scalar field though introduces two
major consequences  worthing to be emphasized.

Besides the fact that the origin $x=0$, $y=0$ is again a fixed
point associated with the vanishing of the scalar field's energy
and, hence, corresponds to the matter domination, the interaction
given by a non-vanishing $r$ has the relevant effect (already
found in the constant $\Gamma$ model) that the stability of the
minima is reinforced and that the maxima become less unstable.
Moreover, the scalar field decay prevents the existence of the
fixed points at $x=\pm 1$, $y=0$, that would correspond to a
behavior completely dominated by the scalar field's kinetic energy
(and which was, therefore,  associated with a stiff behavior in
the $r=0$ case).

The other major effect of the interaction arises when we look for
fixed points with $x^2+y^2<1$ at $\phi \to \infty$. Now, we find
that there are always attracting scaling solutions for potentials
that have an asymptotic exponential behavior, that is, for
potentials for which $W\to {\rm const}$ when $\phi \to
\infty$~\cite{NM 00}. Moreover, this happens independently of the
steepness of the late time exponential behavior which is a
remarkable effect of the present model for the transfer of energy
from the scalar field to the matter.

Indeed the latter solutions are given by the roots of the system of equations
\begin{equation}
(u-1)(u-\frac{a}{b})-ru =0\; , \label{eq_roots}
\end{equation}
and
\begin{equation}
\cos^2{\theta} = \frac{\lambda^2}{6(1+r)^2}\,\xi^2\; , \label{eq_roots2}
\end{equation}
where $u=\xi^2$ and $\theta$ are polar coordinates, $x=\xi \cos \theta$
and $y = \xi \sin \theta$, and where we have defined
\begin{eqnarray}
a &=& \frac{\gamma}{2}\\
b &=& \frac{\lambda^2}{6(1+r)^2}\; ,
\end{eqnarray}
as well as $W_\infty = -\sqrt{3/2} \, \lambda$. It is a simple
matter to conclude that the effect of a non-vanishing $r$ is such
that equations~(\ref{eq_roots}) and (\ref{eq_roots2}) always have
one non-vanishing root within the range of allowed values  for
$\xi$ and for $\cos{\theta}$, and hence there are scaling
solutions regardless of the ratio between $\lambda^2$ and
$3\gamma$. Furthermore linear stability analysis shows that the
scaling solutions are stable. It is  important nevertheless to
remark that although scaling solutions emerge for any ratio of
$\lambda^2/\gamma$, the way that the $\gamma_{eff}$ index
associated with the effective equation of state inducing the
power-law scaling behavior is shifted from the corresponding
$\gamma$ value of the scaling solutions in the absence of decays
depend on $\lambda^2$ being larger or smaller than $\gamma$.

Assuming the potential to be asymptotically given by $V\propto
e^{-\lambda \phi}$, the latter solutions are $a(t) \propto t^A$,
$\phi-\phi_0 = \ln{t^{\pm 2/\lambda}}$, where $A$ is given in
implicit form by
\begin{equation}
3\gamma\left(A-\frac{2}{3\gamma}\right)\left( A-\frac{2}{\lambda^2}(1+r)\right) - \frac{4 r}{\lambda^2}=0 \; .
\label{scal_nopi}
\end{equation}
Notice that we can define $\gamma_{eff}=2/(3A)$.
A linear expansion about $r=0$ in the neighborhood of the scaling
solution (for $\lambda^2\neq 3\gamma$) yields
\begin{equation}
A=\frac{2}{3\gamma}\,\left[ 1+\frac{\frac{2}{\lambda^2}}{\left(\frac{2}{3\gamma}-\frac{2}{\lambda^2}\right)}\,r\right] \; ,
\label{scal_nopi_b}
\end{equation}
when $\lambda^2\neq 3\gamma$. So the decays have the effect of
increasing (resp. decreasing) the scale factor rate of expansion
with regard to the $r=0$ case if $\lambda^2>3\gamma$ (resp.
$\lambda^2<3\gamma$). In particular  we can see that the scaling
behavior can be inflationary, for cases where this would not
happen in the absence of decays. For instance, taking $\gamma=4/3$
and $\lambda^2 >4$, the condition for the scaling solution to be
inflationary is $1+r>\lambda^2/4 >1$. Thus, in this model, the
solutions yield endless power-law inflation even for a modest
scalar field decay, provided that the asymptotic behavior of the
potential is steep enough, i.e., $\lambda^2>3\gamma$ ($>4$ in the
present example).

Naturally, besides the scaling solutions,  there can also be fixed
points corresponding to de Sitter behavior $x=0$, $y=1$, whenever
the scalar field potential exhibits an asymptotic, non-vanishing
constant value. However, when the potential is asymptotically
exponential, that there are no fixed points on the boundary
$x^2+y^2=1$ at $\phi_\infty$ in contrast to what happens in the
$r=0$ case.

From a thermodynamical viewpoint, the above scaling solutions are
particularly interesting. In a universe with two components, it can be
shown~\cite{Zimdahl+Triginer+Pavon 96,Zimdahl+Pavon 01} that the
temperature of each of the components satisfies the equation
\begin{equation}
\frac{\dot{T}_{i}}{T_{i}}= -3\frac{\dot{a}}{a} \left(
1-\frac{\Gamma_{i}}{3H}\right) \frac{\partial p_{i}}{\partial \rho_{i}}+\frac{n_i \dot{s}_i}{\partial \rho_i/\partial T_i}
\; ,
\end{equation}
where $i=1$ or $2$, $n_i$ denotes the number density of particles of the
$i$-species,  $\Gamma_i$ their rate of decay, and $T_i$ the temperature of
this component. In the important case of particle production with constant
entropy per particle, $\dot s_i=0$, we also have $(\rho_1+p_1)\Gamma_1=-(\rho_2+p_2)\Gamma_2$.
Thus, taking the first component to be the matter/radiation fluid and the
second to be the inflaton scalar field, we have
\begin{equation}
\Gamma_\phi=\frac{(\rho+p)}{\dot{\phi}^2}\, \Gamma_{m/r} \propto \Gamma_{m/r}\; .
\end{equation}
As $(\rho+p)/\dot{\phi}^2=\gamma(1-x^2-y^2)/2x^2$ is a constant in the
scaling solutions, $\Gamma_\phi =3rH$, with $r$ a constant,  implies $\Gamma_{m/r}
=3\sigma H$, where $\sigma$ is another constant that depends both on $r$
and on the location of the scaling solution. This yields a temperature of
the matter/radiation  component evolving as a power-law
$T\propto a^{-3(\gamma-1)(1-\sigma)}$. Thus for $\sigma$ close to 1,
the temperature of the matter/radiation remains quasi-static,
whereas for $\sigma>1$ (resp. $\sigma<1$), it increases (resp. decreases).
Notice also that for $\sigma\simeq 0$, we recover the temperature law
for perfect fluids without dissipative effects. Provided we guarantee
enough inflation, $r$ need not be very large (contrary to what is
usually assumed to facilitate slow-rolling).
Indeed, the temperature of the radiation at the end of
inflation is
\begin{equation}
T_{\rm end}= T_{\rm beginning}\, e^{-N(1-\sigma)}\; ,
\end{equation}
where $N$ is the number of e-foldings. Thus, a value of sigma
lower but sufficiently close to $1$ has the potential to avoid a
serious decrease of the temperature of the universe. As
\begin{equation}
\sigma = \frac{3 x_\ast^2}{2(1-x_\ast^2-y_\ast^2)}\, r \; ,
\end{equation}
at the scaling solutions, we see that $r$ need not be very large to ensure that $\sigma \sim 1$.

Trivial scaling solutions generalizing those given Eq.~(\ref{tri_ssol_1}) in the $r=0$ case,
arise in these models for scalar field potentials of
the form
\begin{equation}
V(\phi) =A^2 \,\left(1-\frac{n}{m} \right)^2\,\left( \frac{6(1+r)-n}{2n}\right)\,
\left(\frac{\phi}{A}\right)^\varpi
\label{tri_ssol_3} \; ,
\end{equation}
where $\varpi$ is still given by Eq.~(\ref{tri_ssol_2}) and
$A=A_{r=0}/\sqrt{1+r}$. For the potential to be positive one also
requires $0<m<n<6(1+r)$. The only difference with respect to the
$r=0$ case lies in the dependence on $r$ of the constant factor
multitplying $\phi^\varpi$, and translates the fact that there is
now a different distribution of the scalar field energy density
between its kinetic and potential parts. Indeed, we have that
\begin{equation}
V(\phi) = \left[\frac{6(1+r)}{n}-1\right]\, \dot{\phi}^2
\end{equation}
which shows that the extra damping of the kinetic energy part when
$r\neq 0$, as expected. As in the $r=0$ case, the possible
emergence of the trivial scaling behavior in association with
monomial potentials (that might or not be part of double wells)
happens when the matter fluid is already dominating and is thus of
a lesser importance in the context of warm inflation.

\section{Warm inflation with bulk viscosity} \label{sec_bulkvisc}

One of the main purposes of the present work is to assess the
implications for warm inflation of the presence of a viscous
pressure, $\Pi$, in the matter component, so that the total fluid
pressure is $p=(\gamma-1)\rho+\Pi$. We may assume the expression
$\Pi=-3\zeta H$ which, albeit some causality caveats, is the
simplest one may think of and has been widely considered in the
literature ~\cite{Murphy 73,Weinberg 71,Belinski+Khalatnikov 75}.
If the mixture of massive particles and radiation is taken as a
radiative fluid, it is admissible to adopt $\zeta \propto
\rho_\gamma \tau$, where $\tau$ denotes the relaxation time of the
dissipative process. For the hydrodynamic approach to apply the
condition $t_{col}H<1$ should be fulfilled. Since $\tau \propto
t_{col}$ (a reasonable assumption) and the most obvious time
parameter in this description is $H^{-1}$, one concludes that
$\Pi\simeq -\beta\, \rho_\gamma$, with $0<\beta<1$.

This modifies the field equations (\ref{eq_Ray}) and (\ref{eq_cons})
which now read
\begin{eqnarray}
\dot{H}&=& -\frac{\dot{\phi}^2+\gamma \rho +\Pi}{2} \label{dotH_Pi}\\
\dot{\rho} &=& -3 \left(\gamma+\frac{\Pi}{\rho}\right)\,H\, \rho +
\Gamma \dot{\phi}^2 \; ,
\end{eqnarray}
while equations (\ref{eq_Fried}) and (\ref{eq_KG}) remain in place
and thus the dynamical system is now
\begin{eqnarray}
x' &=& x \,\left[ Q-3 (1+r)\right]- W(\phi) \, y^2,
\label{x'_3} \\
y' &=& \left[Q+W(\phi)\, x\right] y\; ,
\label{y'_3}\\
\chi' &=&  \chi\, \left[ \left(\frac{\Pi'}{\Pi}\right) + 2Q \right]\; , \label{chi'_3} \\
r'  &=& r Q \label{r'_3}\\
\phi' &=& \sqrt{6} \, x \; , \label{vphi'_3}
\end{eqnarray}
where
\begin{equation}
\chi= \Pi/(3H^2) \, ,
\label{chi}
\end{equation}
and
\begin{equation}
Q =\frac{3}{2}\,\left[2x^2+\gamma\,
(1-x^2-y^2)+\chi\right]  \; .
\end{equation}

Apart from raising the order of the system, the main difference
with regard to the previous cases lies in the modification
introduced in $Q$. In fact the bulk viscosity term contributes an
additional term to it, and this changes the properties of some of
the fixed points of the warm inflation model. These implications
naturally depend on the functional form of $\Pi$, and next we
consider some specific choices.

\subsection{Warm inflation with bulk viscosity $\Pi=-3\zeta H$}\label{Subs_PiproptoH}

Our first choice is the ``classical'' assumption already
mentioned, $\Pi= -3\zeta H$, where $\zeta$ is a positive constant
ensuring that the second law of thermodynamics holds. We also
assume that $\Gamma_\phi$ is constant as in
Subsection~\ref{Sec_Gammaconst}.

Given the definition of $\chi$ in the dynamical system
(\ref{x'_3}--\ref{vphi'_3}), we see that $\chi= - \zeta/H \propto
r$. Therefore, defining the constant $\bar{r} =-\Gamma/3\zeta$ so
that we have $r=\bar{r} \chi$, not only Eq.~(\ref{chi'_3})
considerably simplifies, but also we do not need the $r$ equation
(\ref{r'_3}). The resulting dynamical system is
\begin{eqnarray}
x' &=& x \,\left[ Q-3(1+\bar{r}\,\chi )\right]- W(\phi) \, y^2,
\label{x'_4} \\
y' &=& \left[Q+W(\phi)\, x\right] y\; ,
\label{y'_4}\\
\chi' &=&  \chi\, Q \; ,
\label{chi'_4} \\
\phi' &=& \sqrt{6} \, x \; .
\label{vphi'_4}
\end{eqnarray}

We see from Eq.~(\ref{chi'_4}) that this system has fixed points
with vanishing bulk viscosity, $\chi=0$, which send us back to the
cases already studied in  section \ref{sec_dyn_warminf}. The novel
situations, however, arise when the fixed points occur with
$\chi\neq 0$, which requires $Q=0$ (meaning that $H=H_\ast$ is
constant).

For finite values of $\phi$, say at $\phi_\ast$, the fixed points
are defined by  $x_\ast=0$, $W(\phi_\ast)\, y_\ast^2=0$. So we
have a fixed point at  $x_\ast=0$, $y_\ast=0$, $\chi=\chi_\ast$,
$\phi=\phi_\ast$, where $\chi_\ast$ is given by
\begin{equation}
\chi_\ast = -\gamma \; .
\end{equation}
Being associated with the vanishing of both $\dot{\phi}$ and
$V(\phi)$, this is, remarkably, a matter dominated de Sitter
solution.  It is stable if the potential has a vanishing minimum.
Alternatively, we have a line of fixed points given by $x_\ast=0$
and by $\gamma(1-y_\ast^2)=-\chi_\ast$, in accordance to the
$Q_\ast=0$ condition. This solution corresponds again to a de
Sitter exponential behavior and has the remarkable feature that
the energy densities of the scalar field and of the matter remain
in a fixed proportion. It arises in association with an extremum
of the potential $V(\phi)$ and its stability depends on the
extremum being a maximum or a minimum, the maximum being unstable
and  the minimum stable. The presence of $r=\bar{r}\chi $ will
again contribute to render the minima more stable and the maxima
less unstable, as already found in the study of the cases devoid
of viscous pressure.

It is appropriate to emphasize that, for these classes of fixed
points, the matter energy density is not depleted by the
inflationary behavior. This is due, of course, to the well-known
fact that the bulk viscosity contributes a negative pressure and
induces inflationary behavior.

Now, regarding the fixed points arising at $\phi \to \infty$, the
situation is similar to that considered in
Subsection~\ref{Sec_Gammaconst}. In fact the functional forms
adopted both by $\Gamma$ and $\Pi$ prevent the existence of
scaling solutions. The only fixed points associated with the
asymptotic behavior of $V(\phi)$ are analogous to those at finite
$\phi$. If $V(\phi)$ has a vanishing asymptotic value, we have
again both the $x=0$ and $y=0$, $\chi=-\gamma$ de Sitter solutions
dominated by matter, and when $V(\phi)$ exhibits a non-vanishing
asymptotic value at infinity, we have the $x=0$, $\chi=
-\gamma(1-y^2)$ de Sitter solutions characterized by a constant
ratio between the energies of matter and of the scalar field.
Incidentally, one may remark that now  the parameter $r$ that
represents the decay of the scalar field takes a fixed value
determined by the $r=\bar{r}\chi$ relation at the fixed points.

We conclude this Section commenting that if we were to assume the
type of decay considered in Subsection~\ref{Sect_rproptoH}, i.e.,
a constant $r$, the features of the dynamical system would be the
same as in the case just considered. This means that the
remarkable effect we found there that attractor scaling solutions
would always emerge when the potential has an asymptotic
exponential behavior is destroyed by the addition of bulk
viscosity of the type $\Pi=-3\zeta H$.

\subsection{Warm inflation with general bulk viscosity and decay terms}
\label{Subs_Pigeneral}

We now extend our previous analysis to ascertain the implications
of more general functional dependences of both $\Gamma_\phi$ and
$\Pi$.

We assume, quite generally, $\Gamma_\phi=%
\tilde\Gamma(\phi)\,H^\delta$ and $\Pi = -3\zeta\,\rho^\alpha\,
H^\beta$, where $\delta > 0 $, $\zeta$, $\alpha$ and $\beta$ are
constants and moreover $2\alpha+\beta-2=0$ on dimensional grounds.
The latter condition on the parameters $\alpha$ and $\beta$
implies that the dimensionless variable $\chi$ becomes
$\chi=-3^{\alpha}\zeta\, (\rho/3H^2)^\alpha$ and, hence, it
reduces to $\chi=-3^{\alpha}\zeta\, (1-x^2-y^2)^\alpha$, since, as
previously, we still have $\rho/3H^2=1-x^2-y^2$. The $\delta$ free
parameter controls how fast   $\Gamma_\phi$ decreases with
decreasing $H$ during slow-roll inflation (see
Eq.(\ref{dotH_Pi})). Moreover, for $\delta>1$, $r$ decreases with
decreasing $H$, and that for $\delta<1$ it increases.

The dynamical system is 4-dimensional and reads
\begin{eqnarray}
x' &=& x \,\left( Q-3 (1+r)\right)- W(\phi) \, y^2, \label{x'_5} \\
y' &=& \left(Q+W(\phi)\, x\right) y\; , \label{y'_5}\\
r'  &=& r \left[ \sqrt{6}\left(\frac{\partial_\phi\tilde\Gamma}{\tilde\Gamma}\right)\, x + Q (1-\delta) \right] \label{r'_5}\\
\phi' &=& \sqrt{6} \, x \; , \label{vphi'_5}
\end{eqnarray}
where $Q$ is now given by
\begin{equation}
Q =\frac{3}{2}\,\left[2x^2+\gamma\,
(1-x^2-y^2)- 3^{\alpha}\zeta\, (1-x^2-y^2)^\alpha\right] \label{Sigma_5} \; .
\end{equation}
In what follows it seems reasonable to further assume that
$0<\alpha<1$ so that $\beta>0$ which amounts to having a bulk
viscosity pressure whose importance diminishes with the expansion
and with the dilution of matter.

Inspection  of Eqs.~(\ref{x'_5}--\ref{Sigma_5}) shows that it
becomes possible to avoid the restrictive $Q=0$ condition
previously found in the $\Gamma=\;{\rm constant}$ and $\Pi =
-3\zeta\,H$ models that implied a de Sitter behavior at the fixed
points. We have now a wider range of possibilities (our results
for the scaling solutions are summarised in Table~\ref{table3}).

At finite values of $\phi$ we find a line of fixed points
$x=y=r=0$, and another line of fixed points characterized by
$x=0$, $\phi=\phi_0$, $(1-y^2)^{1-\alpha}=3^{\alpha}\zeta/\gamma$
and any value of $r$. In the latter case $\phi_0$ is the value of
$\phi$ at an extremum of $V(\phi)$, i.e., where $W(\phi_0)=0$, and
in order to guarantee that $y^2\leq 1$ we require that
\begin{equation}
3^{\alpha}\zeta <\gamma \label{gen_cond1}\; .
\end{equation}
(Notice that this amounts to having $\rho+p+\Pi > 0$, hence
ensuring $\dot\rho<0$, regardless of the ratio $\rho/(3H^2)$. It
is, thus, a condition akin to the usual weak energy condition).

The linear stability analysis shows that the singular points
$x=y=r=0$ corresponding to matter domination are unstable. In fact
the eigenvalues are
\begin{eqnarray}
\lambda_\phi &=& 0 \\
\lambda_r &=& \frac{3}{2}\,(1-\delta)\,(\gamma-3^{\alpha}\zeta)  \\
\lambda_y &=& \frac{3}{2}\,(\gamma-3^{\alpha}\zeta)>0 \\
\lambda_x &=& 3\left(\frac{\gamma-3^{\alpha}\zeta}{2}-(1+r)\right) \, ,
\end{eqnarray}
and we see that $\lambda_y$ is positive.

Regarding the  line of singular points with $r\neq 0$, linear
stability analysis reveals that, besides the vanishing eigenvalue
associated with $r$ ($\lambda_r=0$), the sability is once more
determined by the nature of the extremum of $V(\phi)$. Indeed, the
eigenvalue corresponding to $y$ is
\begin{equation}
\lambda_y  = 2\gamma\,y_\ast^2\,(\alpha-1)<0 \; , \qquad {\rm when}\;\; \alpha<1 \, ,
\end{equation}
where $y_\ast$ is a solution of $(1-y_\ast^2)^{1-\alpha}=3^{\alpha}\zeta/\gamma$,
and the eigenvalues along $x$ and $\phi$ are given by
\begin{equation}
\lambda_{x,\phi} = - \frac{3(1+r)}{2}\pm \frac{1}{2}\,\sqrt{9(1+r)^2-4\sqrt{6}\,y_\ast^2\,W'(\phi_0)} \, .
\label{eigenv_x5}
\end{equation}
We see from the latter equation that $\lambda_{x,\phi}>0$ requires
that $W'(\phi_\ast)<0$, that is a maximum at $V(\phi_\ast)$.
Otherwise, in the case of a minimum of $V(\phi)$, we have either a
stable node  (when $0<W'(\phi_0)<9(1+r)^2/(4\sqrt{6}\,y_\ast^2)$)
or a stable sink (when $W'(\phi_0)>9(1+r)^2/(4\sqrt{6}\,y_\ast^2)>%
0$).

As is well-known, some authors have resorted to the cooperative
action of many scalar fields -the so-called``c inflation"-, both
in cool inflation \cite{assisted} and in warm inflation
\cite{synergistic}, to get a sufficiently flat effective potential
capable of driving power-law accelerated expansion. Here we note
that this can be achieved with just a single inflationary field
provided the dissipative bulk viscosity is not ignored. Indeed,
this can be seen from from Eq.~(\ref{eigenv_x5}). An increase in
the parameter $r$ as well as a decrease of $y_\ast$ induce an
effective reduction of the steepness of the potential at the
maxima and a greater stability of the minima. Since $y_\ast^2 = 1-
(3^{\alpha}\zeta/\gamma)^{1/(1-\alpha)}$, a decrease in $y_\ast$
translates an increase in $\zeta$ within the admissible range
($3^{\alpha}\zeta/\gamma<1$). The net effect is that the system
spends a longer time in the neighborhood of a fixed point
associated with a  maximum of the potential (alternatively, the
minima become more stable). This is helpful for setting the
conditions for slow-roll inflation. In fact, this alleviates the
need for a large rate of decay of the scalar field. We just need
$\left[3(1+r)/y_\ast\right]^2>1$.

At $\phi\to \infty$, labelled $\phi_\infty$, Eq.~(\ref{r'_5})
shows that we may have the usual fixed points corresponding to a
non-vanishing, flat asymptotic asymptotic behavior of the
potential (late-time approach to a cosmological constant) if
$Q=0$, $\Gamma'/\Gamma=0$ and $\Pi'/\Pi=0$ simultaneously.
However, we also find asymptotic scaling behavior in the case
$W(\phi)$ approaches an exponential behavior
($W(\phi_\infty)=-\sqrt{3/2}\,\lambda$, with constant
$\lambda>0$), provided
\begin{equation}
W(\phi) = \frac{\sqrt{6}}{1-\delta}\,\left(\frac{\partial_\phi \tilde\Gamma}{\tilde\Gamma}\right)\; ,
\end{equation}
which amounts to having, at $\phi \to \infty$, $\displaystyle
\tilde\Gamma(\phi)\propto
\left(V(\phi)\right)^{\frac{1-\delta}{2}}$ and $\tilde\Gamma$ must
be asymptotically exponential. Notice that for $\delta=1$ we
recover the $\Gamma \propto H$ rate of decay considered in
section~(\ref{Sect_rproptoH}). The scaling solutions are then
characterized in polar coordinates, $x=\xi\, \cos{\theta}$,
$y=\xi\, \sin{\theta}$, by
\begin{equation}
\cos{\theta_\ast} = \frac{\lambda}{\sqrt{6}(1+r)}\,\sqrt{u}\; , \label{sp_theta_gen5}
\end{equation}
where $u\equiv \xi^2$ is a root of the equation
\begin{equation}
(1-u)\,(a-b u) - r b u - \frac{3^{\alpha}\zeta}{2}\,(1-u)^\alpha =0 \; . \label{sp_rho_gen5}
\end{equation}
The quantities $a$ and $b$ were defined above.

We see from Eqs.~(\ref{sp_theta_gen5}) and (\ref{sp_rho_gen5})
that there is always one (and only one) scaling solution, provided
the condition~(\ref{gen_cond1}) holds. (Notice that this was
precisely the condition that was required for the existence of
fixed points at finite $\phi$). Indeed, the first two terms of Eq.
~(\ref{sp_rho_gen5}) are a second-order polynomial $P_2(u)$ with
$P_2(0)>0$ and $P_2(1)<0$, so that it has one, and only one zero,
in that interval $(0,1)$. Thus the addition (subtraction) of the
$\frac{3^{\alpha}\zeta}{2}\,(1-u)^\alpha$ has the net effect of
making the root of $P_2$ approach the origin $u=0$, and the latter
remains in the $(0,1)$ interval provided ~(\ref{gen_cond1}) is
valid. We also find that the location of the singular points
corresponding to the scaling behavior is now closer to
$x^2+y^2=0$, having a smaller $y_\infty$ value than in the models
without bulk viscosity. Moreover, linear stability analysis
reveals that under the conditions (\ref{gen_cond1}) and $\alpha<1$
the scaling solutions are stable, i.e., are attractors. These
results mean that the late time contribution of the matter
component is enhanced by the viscous pressure. This is a most
convenient feature for the warm inflation scenario, since it
further alleviates the depletion of matter during inflation and
the subsequent need for reheating.

The power law behavior of these solutions is $a\propto t^A$,
$\phi\sim \ln{t^{2/\lambda}}$ with $A$ given in implicit form by
\begin{equation}
\left(3\gamma\,A-2\right)\,\left[A-\frac{2}{\lambda^2}(1+r)\right]=
\frac{4}{\lambda^2}r + 3^{1+\alpha}\zeta
A^{2-\alpha}\,\left[A-\frac{2}{\lambda^2}(1+r)\right]^\alpha \; ,
\end{equation}
where $r$ is here the asymptotic value of this parameter at the
scaling solution (where, $r\propto y_\ast^{\frac{1-\delta}{2}}$).
A linear expansion in both $r$ and $\zeta$ in the neighborhood of
the scaling solution when $r=0$, $\zeta=0$, and $\lambda^2\neq
3\gamma$, yields
\begin{equation}
A(\gamma,\lambda;r,\zeta) \simeq \frac{2}{3\gamma}\,
\left[ 1+\frac{\left(\frac{2}{\lambda^2}\right)}{\frac{2}{3\gamma}-\frac{2}{\lambda^2}}\,r+\frac{3^{1+\alpha}}{2}
\left( \frac{2}{3\gamma}\right)^{2-\alpha}\, \left(\frac{2}{3\gamma}-\frac{2}{\lambda^2}\right)^{\alpha-1}\,\zeta \right] \; .
\end{equation}
Naturally, these equations reduce to the Eqs.~(\ref{scal_nopi})
and (\ref{scal_nopi_b}) when $\delta=1$ and $\zeta=0$. As found in
subsection~\ref{Sect_rproptoH}, it is possible to define in the
same manner a $\gamma_{eff}=2/(3A)$. We see that now it might be
possible to have $\lambda^2<3\gamma$ if $\alpha$ is, for instance,
a rational of the type $\alpha=m/(2n)$ with $m(<2n)$ and $n$
integers. However for $\alpha=m/(2n+1)$, $\lambda^2$ can be both
larger or smaller than $3\gamma$, rspectively, yielding
$\gamma_{eff}$ smaller or larger than $\gamma$. In other words,
larger or smaller values of $A$ with regard to the case without
either decays or bulk viscosity.

\begin{figure}
\begin{center}
\psfig{file=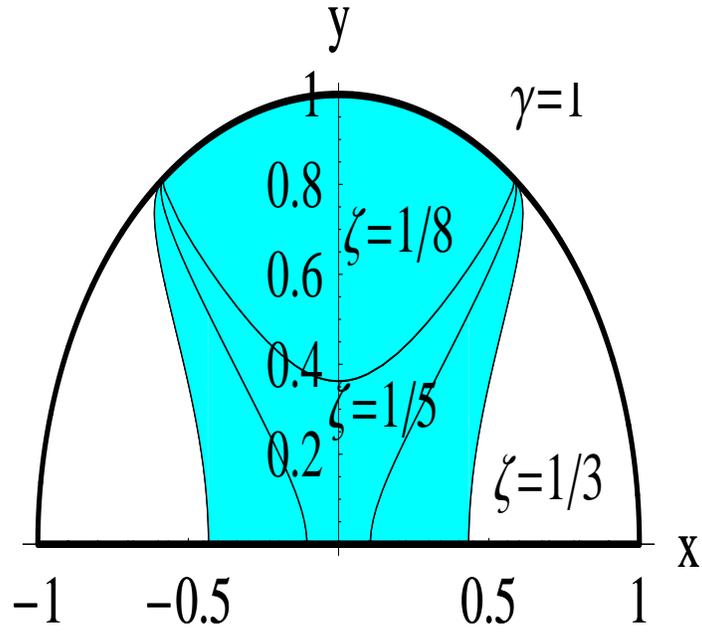,height=12cm,width=12cm}
\end{center}
\caption{Inflationary region of the models with bulk viscosity for
$\gamma=1$. Inflation occurs in the shaded regions between the
border lines and the $x^2+y^2=1$ boundary of the phase space. The
lowest line corresponds to $\zeta=1/3$, the intermediate line to
$\zeta=1/5$ and the uppest line to $\zeta=1/8$. We see that the
size of the inflationary region decreases with  $\zeta$.}
  \label{fig2}
\end{figure}

It is interesting to look at the modifications of the regions of
the phase space that correspond to inflationary behavior arising
from the consideration of the viscous pressure. In  particular, it
is important to assess how they depend on the choice of the
parameters. From Figure~\ref{fig2} we see that the size of the
inflationary region is larger than in the corresponding models
without viscous pressure (models with the same $\gamma$). This was
expected as the bulk viscosity term amounts to a negative
pressure. We also see, in good agreement with this, that the size
of the inflationary region decreases with decreasing $\zeta$, as
the importance of the viscous pressure diminishes.

We conclude this section by briefly commenting on the trivial
scaling solutions. The scalar field potentials that yield trivial
scaling solutions when the matter fluid dominates are still given
by Eq.~(\ref{tri_ssol_3}). The requirement that $r$ be a constant
now translates into
\begin{equation}
\tilde{\Gamma}(\phi) \propto \left[ \left(V(\phi)\right)^{\frac{\delta-1}{2}}\right]^{\frac{2-\varpi}{\varpi}} \; ,
\end{equation}
and in addition we have a consistency condition
\begin{equation}
\frac{1}{3}\left( \frac{m-\gamma}{3\zeta}\right)^{\frac{1}{\alpha-1}} = 1 \; .
\end{equation}
Once again, the inequalities $0<m<n<6(1+r)$ must be satisfied if
the potential is to be positive. Moreover, as in the $\Gamma \propto H$ case, these scaling
solutions happen when matter dominates so that they are not important for warm inflation.

\section{Discussion and conclusions}
In this work we have analyzed the dynamical implications for the
warm inflation scenario of the existence of a viscous pressure in
the matter content of the Universe. The dissipative pressure may
arise either because the fluid in which the inflaton decays may be
treated as a radiative fluid or because the different particles
making up the fluid cool a different rates or because the
particles in which the inflaton decays experience a subsequent
decay in another particles species. We have adopted a
phenomenological approach and have classified the asymptotic
behavior of models associated with possible choices of the
inflaton potential, as well as those arising from various
functional dependences both of the rate of decay of the scalar
field and of the viscous pressure on the matter/radiation
component. In general terms we have considered $\Gamma_\phi=
\Gamma(\phi)\,H^\delta$, where $\delta$ is a constant, and $\Pi=
-3\zeta\,\rho^\alpha\,H^{2(1-\alpha)}$, $\alpha<1$ being a
constant.

Relevant asymptotic regimes arise in association with maxima and
minima of the inflaton potential, at finite $\phi$, and with the
asymptotic exponential behavior of the potential at $\phi_\infty$.
In the latter case, we have found  that the existence of scaling
solutions depends on the form of the decay rate of the scalar
field. Indeed, a necessary and sufficient condition to have
scaling solutions is that the rate of decay $\Gamma_\phi \propto
\Gamma(\phi)H^\delta$ becomes proportional to $H$ and, thus,
$\Gamma(\phi)$ is required to become asymptotically exponential,
as $\Gamma_\infty \propto (V_\infty)^{\frac{\delta-1}{2}}$. In
contrast to the scaling solutions found in the models without
decays (the $r=0$ models), here we find scaling solutions
regardless of the steepness of the potential, that is  for any
combination of $V'/V=-\lambda$ and $\gamma$. However the ratio
between these values defines whether the effective value of the
$\gamma$-index characterising the scaling behavior, and hence the
behavior of the scalar field itself, is larger or smaller than the
corresponding value of $\gamma$ for the models without decays.
Indeed, $\gamma_{eff}<\gamma$ when $\lambda^2>3\gamma$ and,
conversely, $\gamma_{eff}>\gamma$ when $\lambda^2<3\gamma$.

Moreover inflation may be facilitated and is of the power-law
type. On the one hand, inflationary behavior emerges in
association with small values of the $r$ parameter. On the other
hand, the additional presence of bulk viscosity helps in avoiding
a difficulty faced by the warm inflation scenario that was raised
by Yokoyama and Linde~\cite{Yokoyama+Linde 99}. Their argument was
that if, on the one hand, to enhance slow-roll and simultaneously
avoid the depletion of matter, one should have a sufficiently high
rate of decay of the scalar field, on the other hand, this would
make inflation stop earlier, since the transfer of energy from the
scalar field to matter would make the conditions for the
domination of the scalar field cease swiftly.

Overall, the presence of dissipative pressure in the matter
component (which arises on very general physical grounds) lends
strength to the warm inflationary proposal.

\section*{Acknowledgments}
JPM and AN wish to acknowledge the financial support from
``Funda\c c\~ao de Ci\^encia e Tecnologia"  under  the CERN grant
POCTI/FNU/49511/2002 and the C.F.T.C. project POCTI/ISFL/2/618,
and are grateful to J.A.S. Lima for helpful
discussions. DP is grateful to the  Centro de F\'{\i}sica
Te\'orica e Computacional da Universidade de Lisboa for warm
hospitality financial support. This research was partially
supported by the old Spanish Ministry of Science and Technology
under Grants BFM2003-06033 and the ``Direcci\'{o} General de
Recerca de Catalunya" under Grant No. 2001 SGR-00186.

\clearpage


\clearpage
\clearpage


\begin{table}
\caption{The properties of the asymptotic behavior of the
$\Gamma=0$ model. In this table, S stands for stable, U for
unstable, MD for matter dominated, SFD for scalar field dominated,
Min for minimum, Max for maximum, and Scal. Sol. for scaling
solution.}\label{table1}
\begin{ruledtabular}
\begin{tabular}{|c|c|c|c|c|c|c|}
r & $x_\ast$ & $y_\ast$ & $\phi$ & $V(\phi)$ & Eq. state &  Stability \\
\hline
& $0$ & $0$ &  $\phi_\ast$  & $V_\ast=0$ Min($V$)& -- &  S/MD\\
\cline{2-7}
& $0$ & $0$ &  $\phi_\ast$ & $V_\ast=0$& -- &  U/MD\\
\cline{2-7}
 & $0$ & $1$& $\phi_\ast$ & $V_\ast\neq 0$ Min($V$)& $\gamma_\phi=0$ &   S/SFD\\
\cline{2-7}
$0$ & $0$ & $1$& $\phi_\ast$ & $V_\ast\neq 0$ Max($V$) & $\gamma_\phi=0$ &   U/SFD\\
\cline{2-7}
& $0$ & $0$ & $\phi_\infty$ & $V_\infty=0$ & -- & U /MD\\
\cline{2-7}
& $0$ & $1$ & $\phi_\infty$ & $V_\infty \neq 0$ & $\gamma_\phi=0$ & S/SFD\\
\cline{2-7}
& $\pm 1$ & $0$ & $\phi_\infty$ & $V\sim e^{-\lambda\phi}$ & $\gamma_\phi=2$ & U. (saddle)/SFD\\
\cline{2-7}
& $\displaystyle{\lambda/\sqrt{6}}$ & $\sqrt{1-\lambda^2/6}$ & $\phi_\infty$ & $V\sim e^{-\lambda\phi}$ & $\gamma_\phi=\lambda^2/3$ & S (node)/SFD\\
\cline{2-7}
& $\displaystyle{\lambda/\sqrt{6}}$ & $\sqrt{1-\lambda^2/6}$ & $\phi_\infty$ & $V\sim e^{-\lambda\phi}$ & $\gamma_\phi=\lambda^2/3$ & U. (saddle)/SFD\\
\cline{2-7}
& $\displaystyle{\sqrt{\frac{3}{2}}\left(\frac{\gamma}{\lambda}\right)}$ & $\displaystyle{\sqrt{\frac{3(2-\gamma)\gamma}{2\lambda^2}}}$ & $\phi_\infty$ & $V\sim e^{-\lambda\phi}$ & $\gamma$ & S/Scal. Sol., $3\gamma<\lambda^2<\frac{24\gamma^2}{9\gamma-2}$\\
\cline{2-7}
& $\displaystyle{\sqrt{\frac{3}{2}}\left(\frac{\gamma}{\lambda}\right)}$ & $\displaystyle{\sqrt{\frac{3(2-\gamma)\gamma}{2\lambda^2}}}$ & $\phi_\infty$ & $V\sim e^{-\lambda\phi}$ & $\gamma$ & S/Scal. Sol., $\lambda^2>\frac{24\gamma^2}{9\gamma-2}$\\
\end{tabular}
\end{ruledtabular}
\end{table}

\clearpage

\begin{table}
\caption{The properties of the asymptotic behavior of the
$\Gamma\propto H$ model. In this table, S stands for stable, U for
unstable, MD for matter dominated, SFD for scalar field dominated,
Min for minimum, Max for maximum, and Scal. Sol. for scaling
solution. The $\gamma_{eff}$ index is defined by
$\gamma_{eff}=2/(3A)$ where $A$ characterizes the power-law
expansion $a\propto t^A$.}\label{table2}

\begin{tabular}{|c|c|c|c|c|c|}
\hline
$x_\ast$ & $y_\ast$ & $\phi$ & $V(\phi)$ & Eq. state &  Stability \\
\hline
$0$ & $0$ &  $\phi_\ast$  & $V_\ast=0$ Min($V$)& -- &  S/MD\\
\hline
$0$ & $0$ &  $\phi_\ast$ & $V_\ast=0$& -- &  U/MD\\
\hline
 $0$ & $1$& $\phi_\ast$ & $V_\ast\neq 0$ Min($V$)& $\gamma_\phi=0$ &   S/SFD\\
\hline
$0$ & $1$& $\phi_\ast$ & $V_\ast\neq 0$ Max($V$). & $\gamma_\phi=0$ &   U/SFD\\
\hline
$0$ & $0$ & $\phi_\infty$ & $V_\infty=0$ & -- & U /MD\\
\hline
$0$ & $1$ & $\phi_\infty$ & $V_\infty \neq 0$ & $\gamma_\phi=0$ & S/SFD\\
\hline
$0<x_0<1$ & $0<y_0<1$ & $\phi_\infty$ & $V\sim e^{-\lambda\phi}$ & $\gamma_{eff}<\gamma$ & S/Scal. Sol., $\lambda^2>3\gamma$\\
\hline
$0<x_0<1$ & $0<y_0<1$ & $\phi_\infty$ & $V\sim e^{-\lambda\phi}$ & $\gamma_{eff}>\gamma$ & S/Scal. Sol., $\lambda^2<3\gamma$\\
\hline
\end{tabular}
\end{table}

\clearpage

\begin{table}
\caption{The properties of the scaling, asymptotic behavior of the
$\Gamma\propto \Gamma(\phi)H^\delta$, $\Pi=-3\zeta \rho^\alpha
H^{2(1-\alpha)}$ model: In this table S stands for stable and
Scal. Sol. stands for scaling solution. The $\gamma_{eff}$ index
is defined by $\gamma_{eff}=2/(3A)$ where $A$ characterizes the
power-law expansion $a\propto t^A$.}\label{table3}

\begin{tabular}{|c|c|c|c|c|c|}
\hline
$x_\ast$ & $y_\ast$ & $\phi$ & $V(\phi)$ & Eq. state &  Stability \\
\hline
$0<x_0<1$ & $0<y_0<1$ & $\phi_\infty$ & $V\sim e^{-\lambda\phi}$, $\Gamma(\phi)\sim V_\infty^{\frac{1-\delta}{2}}$ & $\gamma_{eff}<\gamma$ & S/Scal. Sol., $\lambda^2>3\gamma$\\
\hline
$0<x_0<1$ & $0<y_0<1$ & $\phi_\infty$ & $V\sim e^{-\lambda\phi}$, $\Gamma(\phi)\sim V_\infty^{\frac{1-\delta}{2}}$  & $\gamma_{eff}>\gamma$ & S/Scal. Sol., $\lambda^2<3\gamma$\\
\hline
\end{tabular}
\end{table}


\begin{thebibliography}{99}
\bibitem{Inflation} E. W. Kolb and M. S. Turner, {\em The Early Universe} (Addison-Wesley,
Reading, Massachussetts, 1990); A. R. Liddle and D. Lyth, {\em
Cosmological Inflation and Large Scale Structure} (Cambridge
University Press, Cambridge, 2000); A. Linde, ``Particle Physics
and Inflationary Cosmology", hep-th/0503203.

\bibitem{Berera Pascos}
A. Berera, ``Dissipative dynamics of inflation, in {\em Particles,
Strings", and Cosmology}, Proceedings of the Eighth International
Conference, eds. P. Frampton and J. Ng (Rinton Press, 2001.,
p.393), hep-ph/0106310.

\bibitem{Kofman et al 97} L. Kofman, A.D. Linde and A.A. Starobinsky, Phys. Rev. D 56:3258 (1997),
hep-ph/9704452.


\bibitem{Bassett+Maartens 99}
B.A. Bassett, D.I. Kaiser and R. Maartens,  Phys. Lett. B, 455:84
(1999).

\bibitem{Charters+Nunes+Mimoso 05a} T. Charters, A. Nunes and J. P. Mimoso,  Phys. Rev. D71:083515 (2005),
hep-ph/0502053.

\bibitem{Berera 95a}
A. Berera, Phys. Rev. Lett., 75:3218 (1995).

\bibitem{Berera 95b}
A. Berera and L.Z. Fang, Phys. Rev. Lett., 74:1912 (1995).
\bibitem{Berera+Ramos 05} A. Berera and R. O. Ramos, Phys. Rev. D 71:023513 (2005).
\bibitem{Hall} L.M. Hall and I.G. Moss, Phys. Rev. D 71:023514
(2005).
\bibitem{Mar} M. Bastero-Gil and A. Berera, Phys. Rev. D 71:063515 (2005).

\bibitem{origin}
A. Berera, Nucl. Phys. B585:666 (2000); A.N. Taylor and A. Berera,
Phys. Rev. D 62:083517 (2000).

\bibitem{Hall+Moss+Berera 03} L.M.H. Hall, I.G. Moss and A. Berera, Phys. Rev. D 69:083525 (2004),
astro-ph/0305015.

\bibitem{Gupta etal 02} S. Gupta et al, Phys. Rev. D 66:043510 (2002).
\bibitem{baryogenesis} R.H. Brandenberger and M. Yamaguchi, Phys.
Rev. D 68:023505 (2003).
\bibitem{landau} L. Landau and E.M. Lifshitz, {\em M\'{e}canique des
Fluides} (MIR, Moscou, 1971); K. Huang, {\em Statistical
Mechanics} (J. Wiley, 1987).

\bibitem{Ya-JDB} Ya. B. Zel'dovich, Sov. Phys. JETP Lett. 12:307 (1970);
J.D. Barrow, Nucl. Phys. B 310:743 (1988).
\bibitem{radiative1} S. Weinberg, {\em Gravitation and
Cosmology}, pp. 51-58 (J. Wiley, N.Y., 1972).
\bibitem{radiative2}
 N. Udey and W. Israel, Mon. Not. R. Astr. Soc. 199:1137 (1982).

\bibitem{radiative3} D. Jou and D. Pav\'on, Astrophys J., 291:447
(1983).
\bibitem{Harris}
S. Harris, {\em An Introduction to the Theory of Boltzmann
Equation} (Holt, Reinhart and Winston, N.Y., 1971); C. Cercignani,
{\em Theory and Applications of the Boltzmann Equation} (Scottish
Academic Press, Edinburgh, 1975).


\bibitem{analogy} W. Zimdahl and D. Pav\'{o}n, Phys. Lett. A
176:57 (1993); W. Zimdahl and D. Pav\'{o}n , Mon. Not. R. Astron.
Soc. 266:872 (1994); W. Zimdahl and D. Pav\'{o}n , Gen. Relativ.
Grav. 26:1259 (1994); W. Zimdahl, Mon. Not. R. Astron. Soc. 280:12
(1996), W. Zimdahl, Phys. Rev. D 53:5483 (1996).

\bibitem{decay} A. Berera and R.O. Ramos, Phys. Lett. B 567:294
(2003); A. Berera and R.O. Ramos, Phys. Rev. D 71:023513 (2005).



\bibitem{Calzetta+Hu 88} E. Calzetta, B.L. Hu, Phys. Rev. D37:2878 (1988).

\bibitem{Yokoyama+Linde 99} J. Yokoyama and A. Linde, Phys. Rev.
D 60:083509 (1999).
\bibitem{Berera+Gleiser+Ramos 98}
A.Berera, M. Gleiser, R.O. Ramos, Phys. Rev. D 58:123508 (1998),
hep-ph/9803394.

\bibitem{Berera+Gleiser+Ramos 99}
A. Berera, M. Gleiser, R. O. Ramos, Phys. Rev. Lett., 83:264-267
(1999), hep-ph/9809583.

\bibitem{Moss 02}
Ian G Moss, Nucl. Phys. B, 631:500 (2002), hep-ph/0103191.

\bibitem{Lawrie 02}
I.D. Lawrie, Phys. Rev. D 66:041702 (2002), hep-ph/0204184.



\bibitem{Oliveira+Ramos 98}
H.P. de Oliveira and R.O. Ramos, Phys. Rev. D 57:741 (1998),
gr-qc/9710093.

\bibitem{Bellini 98}
M. Bellini, Phys. Lett. B, 428:31 (1998).
\bibitem{Maia+Lima 99}
J.M.F. Maia and J.A.S. Lima, Phys. Rev. D 60:101301 (1999),
astro-ph/9910568).


\bibitem{Yokoyama+Sato+Kodama 87}
J. Yokoyama, K. Sato and H. Kodama, Phys. Lett. B, 196:129 (1987).


\bibitem{Billyard+Coley 00} A.P. Billyard and A.A. Coley, Phys. Rev. D 61:083503 (2000),
astro-ph/9908224.
\bibitem{Albrecht et al 82}
A. Albrecht, P.J. Steinhardt, M.S. Turner and F. Wilczek , Phys.
Rev. Lett., 48:1437 (1982).

\bibitem{NM 00} A. Nunes and J.P. Mimoso, Phys. Lett. B 488:423 (2000).
\bibitem{exponential1}
K.A. Olive, Phys. Rep. 190:308 (1990); M.B. Green, J.H. Schwarz,
and E. Witten, {\em Superstring Theory} (Cambridge University
Press, Cambridge, 1978); A. Salam and E. Sezgin, Phys. Lett B
147:47 (1984).
\bibitem{Wetterich 94} C. Wetterich, Astron. Astrophys. 301:321 (1995),
J.J. Haliwell, Phys. Lett B 185:341 (1987).

\bibitem{NMC 01} A. Nunes, J.P. Mimoso and T.C. Charters, Phys. Rev. D 63:083506 (2001).

\bibitem{MNP AIP04} J.P. Mimoso, A. Nunes and D. Pav\'{o}n, ``Scaling Behaviour in Warm Inflation",
in Proceedings of the AIP Conference {\em Phi in the Sky. The
Quest for Cosmological Scalar Fields} (Porto, Portugal, 2004).

\bibitem{Ratra+Peebles 88} B. Ratra and P.J.E. Peebles, Phys. Rev. D 37:3406 (1988).
\bibitem{WCL 93} D. Wands, E.J. Copeland and A.R. Liddle, Ann. N. Y. Acad. Sci., 688:647 (1993).

\bibitem{Zimdahl+Triginer+Pavon 96} W. Zimdahl, J. Triginer and D. Pav\'{o}n, Phys. Rev. D 54:6101 (1996).

\bibitem{Zimdahl+Pavon 01}
W. Zimdahl and D. Pav\'{o}n, Gen. Rel. Grav., 33:791 (2001),
astro-ph/0005352.
\bibitem{Lima+Carrillo} J.A.S. Lima and J.A. Espichan Carrillo,
``Thermodynamic approach to warm inflation", astro-ph/0201168.


\bibitem{Zlatev et al 99} I. Zlatev, L. Wang and P. J. Steinhardt, Phys. Rev. D59:123504 (1999).
\bibitem{Liddle+Scherrer 99} A. R. Liddle and R. J. Scherrer, Phys. Rev. D59:023509 (1999).

\bibitem{Ferreira+Joyce 97} P.G. Ferreira and M. Joyce, Phys. Rev. Lett., 79:4740 (1997).
\bibitem{CLW 98} E.J. Copeland, A.R. Liddle and D. Wands, Phys. Rev. D 57 (1998) 4686.
\bibitem{Wetterich 88} C. Wetterich, Nucl. Phys. B, 302:668 (1988) .

\bibitem{Belinski et al 85} V.A. Belinski, L.P.  Grischuk, Ya. B. Zeldovich and
I.M. Khalatnikov, Sov. Phys. JETP, 63:195 (1985).
\bibitem{Murphy 73} G. Murphy, Phys. Rev. D 8:423 (1973).
\bibitem{Weinberg 71} S. Weinberg, Astrophys. J., 168:175 (1971).
\bibitem{Belinski+Khalatnikov 75} V.A. Belinski and I.M. Khalatnikov, Sov. Phys. JETP, 62:195 (1978).
\bibitem{assisted}
A.R. Liddle, A. Mazumdar, and F.E. Schunck, Phys. Rev. D
58:0661301 (1998); A.A. Coley and R.J. van den Hoogen, Phys. Rev.
D 62:023517 (2002).
\bibitem{synergistic} L.P. Chimento, A.S. Jakubi, D. Pav\'{o}n and N. Zuccal\'{a},
Phys. Rev. D 65:083510 (2002).
\end{thebibliography}
\end{document}